\documentclass[12pt]{iopart}

\usepackage{graphicx}

\begin{document}

\title[Excitation functions of baryon anomaly and freeze-out properties at RHIC-PHENIX]{Excitation functions of baryon anomaly and freeze-out properties at RHIC-PHENIX}

\author{T Chujo for the PHENIX\footnote{For the full list of PHENIX authors and acknowledgments, see Appendix
'Collaborations' of this volume.} Collaboration}

\address{University of Tsukuba, Institute of Physics, 
 1-1-1 Tennodai, Tsukuba Ibaraki 305-8571, Japan}
\ead{chujo@sakura.cc.tsukuba.ac.jp}

\begin{abstract}
The intermediate $p_T$ region (2 - 5 GeV/$c$) in central Au+Au collisions at RHIC has a rich physics 
content. The (anti)proton to pion ratio at the intermediate $p_T$ gives us a powerful tool to investigate 
the bulk properties of the hot and dense matter created at RHIC and their hadronization processes. We present 
the preliminary results of identified charged hadron spectra at the lower beam energies at RHIC.
The excitation function of (anti)proton to pion ratios from SPS to RHIC are shown. We also discuss
the onset of the baryon enhancement at the high energy heavy ion collisions. 
\end{abstract}


\section{Introduction}
\label{sec:intro}
One of the most surprising observations at the Relativistic Heavy Ion Collider (RHIC) is a particle type 
dependence of the nuclear modification factors $R_{AA}$ at the intermediate transverse momentum (2 $\le$ $p_T$ 
$\le$ 5 GeV/$c$)~\cite{PPG026,PPG015}.  
In Au+Au central collisions at $\sqrt{s_{NN}} = 200 $ GeV, yields for mesons are largely suppressed~\cite{PPG014} 
with respect to the yields in proton-proton collisions at the intermediate $p_T$, while those for baryons 
are not suppressed. The Cronin effect~\cite{cronin1975,antreasyan1979} alone does not explain the observed 
large baryon yield in central Au+Au collisions~\cite{PPG030}. This phenomena is called "Baryon Anomaly 
(or Enhancement) at RHIC". To explain the data, many theoretical models have been proposed. Among those,
a quark recombination process~\cite{recombi} is now believed to be one of the dominant hadronization 
processes at the intermediate $p_T$ in central Au+Au collisions at RHIC, with the support by 
the experimental data of the nuclear modification factor and the elliptic flow for 
$\phi$ meson~\cite{phi_raa,phi_v2}. Now the key questions are the onset effect of baryon enhancement 
and how it evolves as a function of beam energy. In order to answer these questions, the lower energy 
beam data in Au+Au/Cu+Cu/p+p collisions were taken during the successive RHIC runs from 2004 to 2006 
by the PHENIX experiment. In this paper, we present the preliminary results of $p_T$ spectra 
for identified charged particles ($\pi^{\pm}/p/\overline{p}$) in Au+Au/Cu+Cu/p+p collisions at 
$\sqrt{s_{NN}}$ = 62.4 GeV and Cu+Cu collisions at 22.5 GeV from PHENIX. The centrality and 
beam energy dependences of $p(\overline{p})/\pi$ ratio are presented. We also discuss a possible 
onset effect of the baryon enhancement at RHIC energy.

\section{Data analysis and results}
\label{sec:result}

\begin{figure}
\resizebox{\textwidth}{!}{
  \includegraphics{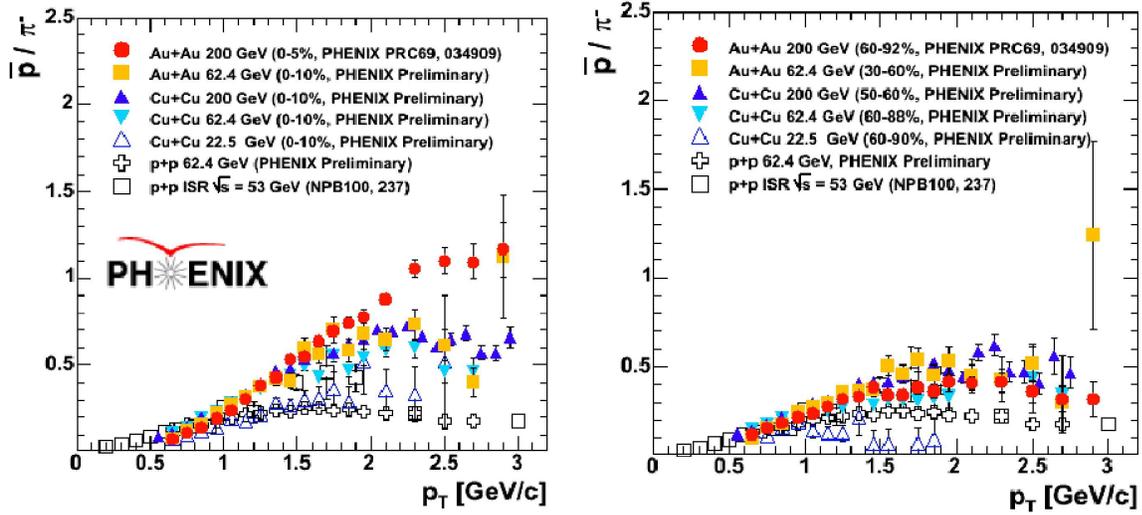}
}
\caption{Compilation of $\overline{p}/\pi^{-}$ ratios as a function of $p_T$ for central
(left) and peripheral (right) collisions.}
\label{fig:ppi_ratio_all} 
\end{figure}

Figure~\ref{fig:ppi_ratio_all} shows the compilation of $\overline{p}/\pi^{-}$ 
ratios as a function of $p_T$, which includes the PHENIX data points in Au+Au/Cu+Cu at 
$\sqrt{s_{NN}} =$ 200 GeV~\cite{PPG026,konno}, Au+Au/Cu+Cu/p+p at 62.4 GeV, and Cu+Cu at 22.5 GeV. 
The ISR data in p+p at $\sqrt{s} =$53 GeV~\cite{Alper75} is also shown for the comparison, 
and it agrees with the newly measured $\overline{p}/\pi$ ratio in p+p at 62.4 GeV.
No weak decay feeddown correction is applied for proton and antiproton yields. 
More details on the data analysis method can be found in~\cite{HQ06_proc}.
For the central collisions (Fig.~\ref{fig:ppi_ratio_all}, left panel), the $\overline{p}/\pi^{-}$ 
ratios at the intermediate $p_T$ in Au+Au/Cu+Cu at 200/62.4 GeV are larger than the values 
in p+p, except for the central Cu+Cu collisions at 22.5 GeV, which is consistent with the p+p values. 
For the peripheral collisions (Fig.~\ref{fig:ppi_ratio_all}, right panel), all of the ratios are 
converging on the ratios in p+p collisions. 

\begin{figure}
\resizebox{0.85\textwidth}{!}{
  \includegraphics{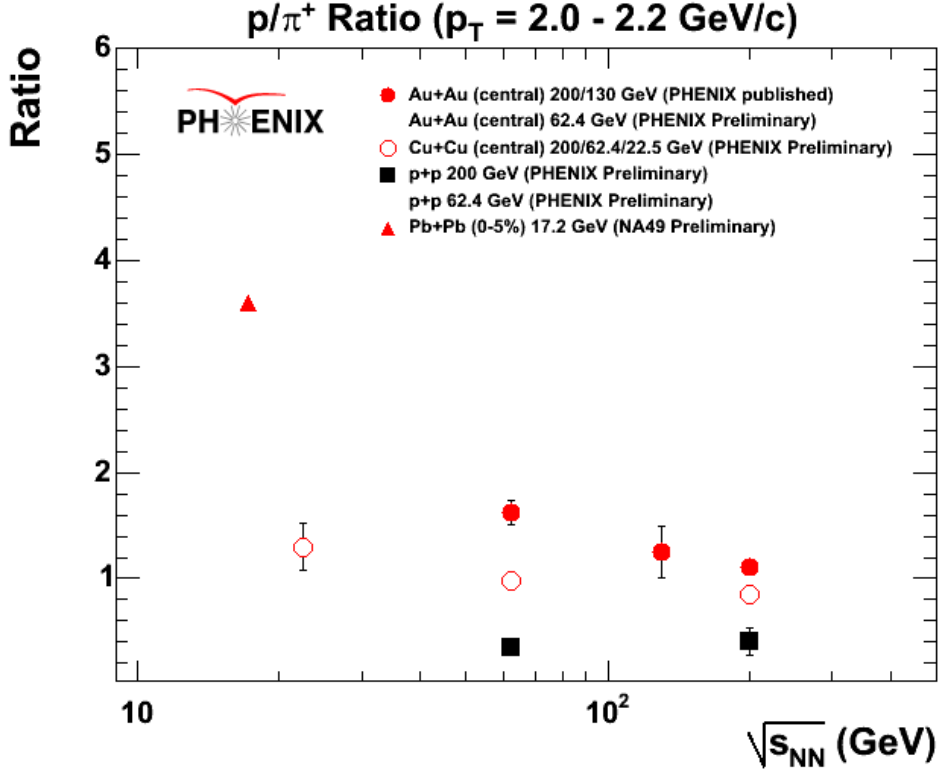}
}
\caption{Beam energy dependence of $p/\pi^{+}$ ratio for the central collisions 
at the intermediate $p_T$ (2.0 - 2.2 GeV/$c$). }
\label{fig:ppi_beam_dep_pos} 
\end{figure}

\begin{figure}
\resizebox{0.85\textwidth}{!}{
  \includegraphics{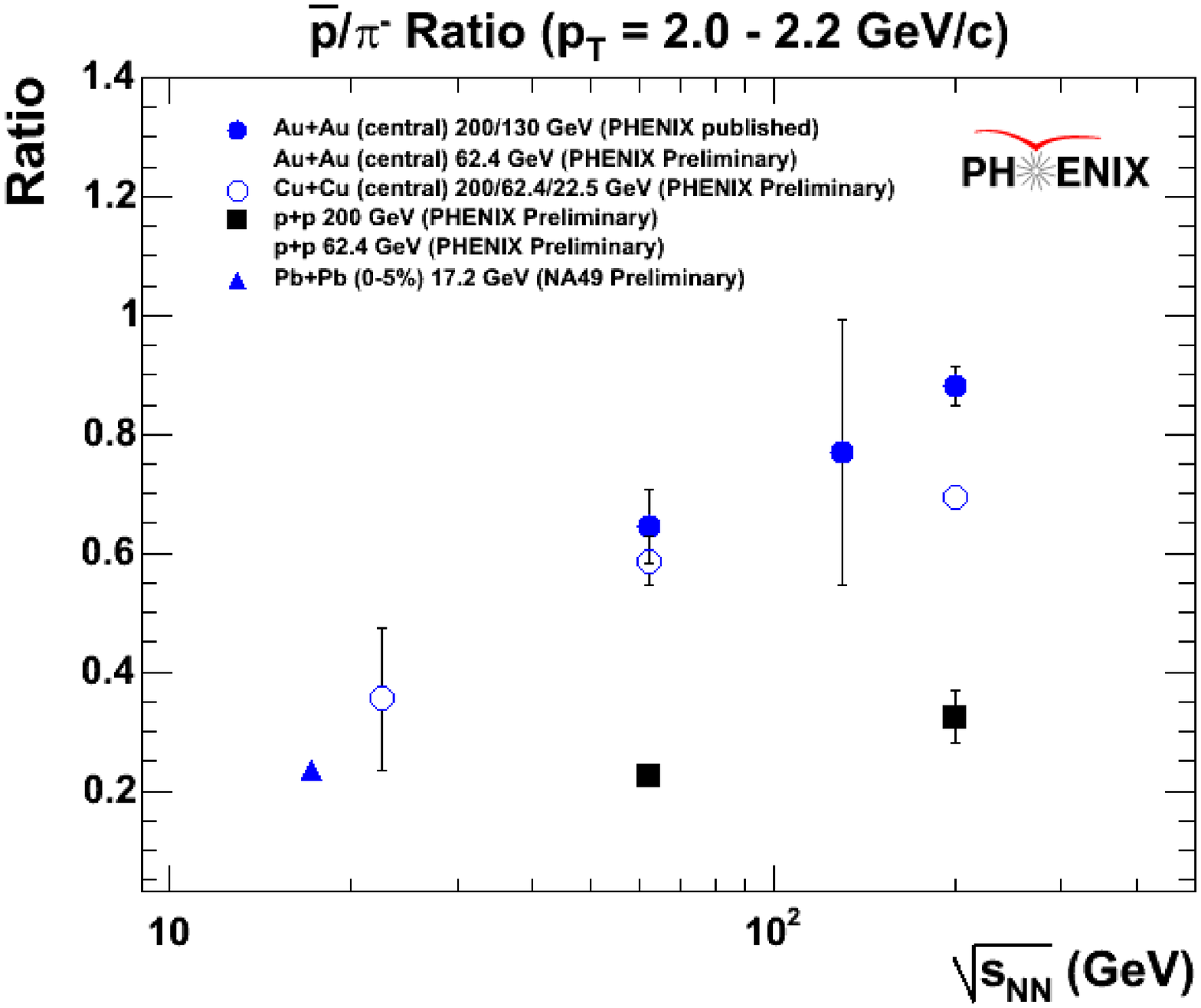}
}
\caption{Beam energy dependence of $\overline{p}/\pi^{-}$ ratio for the central 
collisions at the intermediate $p_T$ (2.0 - 2.2 GeV/$c$).}
\label{fig:ppi_beam_dep_neg} 
\end{figure}

Figure~\ref{fig:ppi_beam_dep_pos} and Figure~\ref{fig:ppi_beam_dep_neg} show the $\sqrt{s_{NN}}$ 
dependence of $p/\pi^{+}$ and $\overline{p}/\pi^{-}$ ratios respectively, from SPS $\sqrt{s_{NN}} =$ 
17.3 GeV~\cite{NA49} to the top RHIC energy 200 GeV in Au+Au/Cu+Cu for the central collisions 
at the intermediate $p_T$ (2.0 - 2.2 GeV/$c$). The data points for p+p are also shown in these plots.
In general, $p/\pi^{+}$ ($\overline{p}/\pi^{-}$) ratio decreases (increases) as a function 
of $\sqrt{s_{NN}}$, respectively. The larger $p/\pi^{+}$ ratios for the lower beam energies 
can be understood by the influence of the incoming nucleons from the beams (a baryon transport
at the midrapidity). On the other hand, $\overline{p}/\pi^{-}$ ratio can be used as a measure
for the baryon enhancement, because antiprotons are ``produced particles''. 
As shown in Figure~\ref{fig:ppi_beam_dep_neg}, the $\overline{p}/\pi^{-}$ ratio in Cu+Cu 22.5 GeV 
is consistent with both p+p measurements and the SPS Pb+Pb central collision data. It suggests 
there is no baryon enhancement at 22.5 GeV in Cu+Cu collisions. An onset effect of baryon 
enhancement could be seen in between $\sqrt{s_{NN}} = $ 22.5 to 62.4 GeV. To conclude, the high 
statistics data with the heavier collisions system like Au+Au at around 22.5 GeV is necessary 
in the future RHIC run, as well as the more extensive beam energy scan in the RHIC-II program.

\section{Summary}
\label{sec:summary}

In summary, we have measured $p/\pi^{+}$ and $\overline{p}/\pi^{-}$ ratios in Au+Au, 
Cu+Cu, and p+p collisions at $\sqrt{s_{NN}} =$ 62.4 GeV, and Cu+Cu at 22.5 GeV.
In the central Cu+Cu collisions at 22.5 GeV, $\overline{p}/\pi^{-}$ ratio at the 
intermediate $p_T$ is consistent with the values in both p+p collisions and Pb+Pb 
central collision, and it suggests that there is no strong baryon enhancement at 22.5 GeV Cu+Cu 
collisions and the onset of baryon enhancement might exist in between $\sqrt{s_{NN}} =$ 
22.5 to 62.4 GeV. These $p/\pi^{+}$ and $\overline{p}/\pi^{-}$ ratios may give 
a further constraint to the hadronization process in the recombination models 
at the beam energies from SPS to RHIC. 


\end{document}